\documentclass[twoside]{dis07}
\usepackage[latin1]{inputenc}
\usepackage[dvips]{graphicx,epsfig,color}
\usepackage{wrapfig,rotating}
\usepackage{amssymb,amsmath,array}

\begin{document}
\title{Dijet Azimuthal Correlations in QCD Hard Processes}

\author{Yazid Delenda
%
\thanks{Work done in collaboration with Andrea Banfi and Mrinal Dasgupta.}
%
\vspace{.3cm}\\
%
School of Physics and Astronomy - The University of Manchester \\
Oxford Road, Manchester M13 9PL - U.K.
%
}

\maketitle \vspace{-5.0cm} \hfill MAN/HEP/2007/10 \vspace{4.55cm}

\begin{abstract}
We study the azimuthal correlation distribution for dijet production
in QCD hard processes. This observable is sensitive to soft and/or
collinear emissions in the back-to-back region, giving rise to
single and double logarithms. We provide resummed predictions to NLL
accuracy for both DIS at HERA and hadronic collisions at Tevatron
and perform a NLO matching to NLOJET++ results in the DIS case.
\end{abstract}

\section{Introduction}

Studies of soft gluon radiation and non-perturbative effects in QCD
observables are of vital importance. These studies help us better
understand the dynamics of QCD and enhance the accuracy of
theoretical predictions for measured quantities. In several
instances precision in QCD is limited not just by what powers of
$\alpha_s$ are controlled, but also by the lack of better
understanding of QCD dynamics such as the all-orders behaviour
(embodied in the resummation of large logarithms) and inevitably the
process of hadronisation.

Successful examples of such studies are manifested in event-shape
variables at LEP and HERA. Resummed estimates for these observables,
combined with NLO predictions and corrected for non-perturbative
effects, have been very successful in describing the data
\cite{Dasgupta:2003iq,Dasgupta:2002dc}. Parameters such as the
strong coupling and the effective non-perturbative coupling
\cite{Dokshitzer:1995zt} can then be consistently extracted by
studying distributions and mean values of such observables (see for
instance Ref. \cite{Kluth:2006hn} for a recent review).

Going beyond the case of two hard partons is more challenging in
terms of theory but is also a more stringent test of our
understanding of QCD dynamics. Multi-jet event-shape variables have
been studied (see Refs. \cite{Banfi:2000si,Banfi:DIS2007}). However
for jet-defined quantities, e.g. several dijet distributions, there
are currently very few resummed predictions because of the lack of
theoretical insight to all orders in the presence of a jet
algorithm. Many measurements are already established (see e.g.
\cite{Hansson:2006ht,Abazov:2004hm}) and await comparison to
theoretical estimates.

Effort has recently been devoted to improve the understanding of the
effect of jet algorithms on QCD resummation
\cite{Appleby:2002ke,Appleby:2003sj,Banfi:2005gj,Delenda:2006nf}. A
clustering algorithm has an impact on the resummation of observables
which are defined in a limited region of the phase-space (such as
energy flow outside jets), known as ``non-global'' observables
\cite{Dasgupta:2001sh,Dasgupta:2002bw}. These receive single logs
which could only be resummed numerically in the large $N_c$ limit
for processes involving only two hard partons. It was shown in Ref.
\cite{Appleby:2002ke} that employing a $k_t$ algorithm on the
final-state particles reduces these logarithms in the case where
only two hard partons are present. However the resummation of
jet-defined quantities proved to be non-trivial \cite{Banfi:2005gj}
and the full impact of clustering algorithms on resummation has been
explained more recently in Ref. \cite{Delenda:2006nf}.

With the technique of resummation using a clustering algorithm one
can proceed with studying jet-defined quantities. In the present
work we focus on the dijet azimuthal correlation distribution. We
consider the process of production of two hard jets in DIS or
hadronic collisions. We study the azimuthal correlation defined by
the azimuthal angle between the leading hard jets in the final
state.

The azimuthal angle of a jet is defined by:
\begin{equation}\label{eq:delenda_def}
\phi_{\mathrm{jet}}=\frac{\sum_{i\in \mathrm{jet}}E_{t,i}\phi_i}
{\sum_{i\in \mathrm{jet}}E_{t,i}}\,,
\end{equation}
where the sum runs over all particles inside the jet. The observable
we study has the following approximation in the soft and/or
collinear regime:
\begin{eqnarray}
\Delta\phi &=& \left|\pi-\delta\phi_{\mathrm{jets}}\right|
,\nonumber\\
&=&\left|\sum_i\frac{k_{t,i}}{p_t}\left(\sin\phi_i-\theta_{i1}
\phi_i-\theta_{i2}(\pi-\phi_i)\right)\right|,
\end{eqnarray}
where $p_t$ is the transverse momentum of the outgoing hard partons,
which we assumed to be at azimuths $\phi_1=0$ and $\phi_2=\pi$. Here
$\theta_{ij}=1$ if particle $i$ is clustered to jet $j$ and is zero
otherwise.

The above definition implies that the observable in question is
global. This means that no non-global component is present and the
resummed result to next-to-leading log (NLL) accuracy has no
dependence on the jet algorithm. This is the recombination scheme
used by the H1 collaboration at HERA to measure this observable
\cite{Hansson:2006ht}. However if one employs a recombination scheme
in which the four-momentum of the jet is defined by the addition of
four-momenta of particles in the jet, then our observable becomes
non-global. In this case one would need to calculate the additional
non-global component as well as the dependence on the jet algorithm.
The D\O\ collaboration at Tevatron employed the latter recombination
scheme to measure the observable
\cite{Abazov:2004hm,Zielinski:2006tr}.

We note that in the soft and/or collinear region, i.e. close to the
Born configuration in which the outgoing jets are back-to-back
($\Delta\phi\sim 0$), the distribution receives large logarithms.
This region is also strongly affected by non-perturbative effects.
In the present study we shall report the resummed predictions for
these logarithms to NLL accuracy both in DIS and hadronic collisions
and provide a matching to NLO results obtained from NLOJET++
\cite{Nagy:2001xb} in the DIS case.

\section{Resummation and matching}

The resummed result for the integrated distribution for events with
$\Delta\phi<\Delta$ is given by:
\begin{equation}\label{eq:delenda_result}
\sigma(\Delta)=\int d\mathcal{B} \frac{2}{\pi} \int_{0}^{+\infty}
\frac{db}{b}\sin(\Delta b) \sigma_{\mathcal{B}} (b) e^{-R(b)}\,,
\end{equation}
where we reported the result for the DIS case assuming the azimuths
of the jets are recombined using Eq.~\eqref{eq:delenda_def}. In
Eq.~\eqref{eq:delenda_result} $\sigma_{\mathcal{B}}(b)$ is the Born
cross-section for the production of two hard jets in DIS, containing
parton distribution functions (pdfs) evolved to $\mu_f^2/b^2$, with
$\mu_f$ being the factorisation scale, and $d\mathcal{B}$ is the
corresponding phase-space. The function $R(b)$ is the radiator which
contains the resummed leading and next-to-leading logs.

\begin{figure}
\epsfig{file=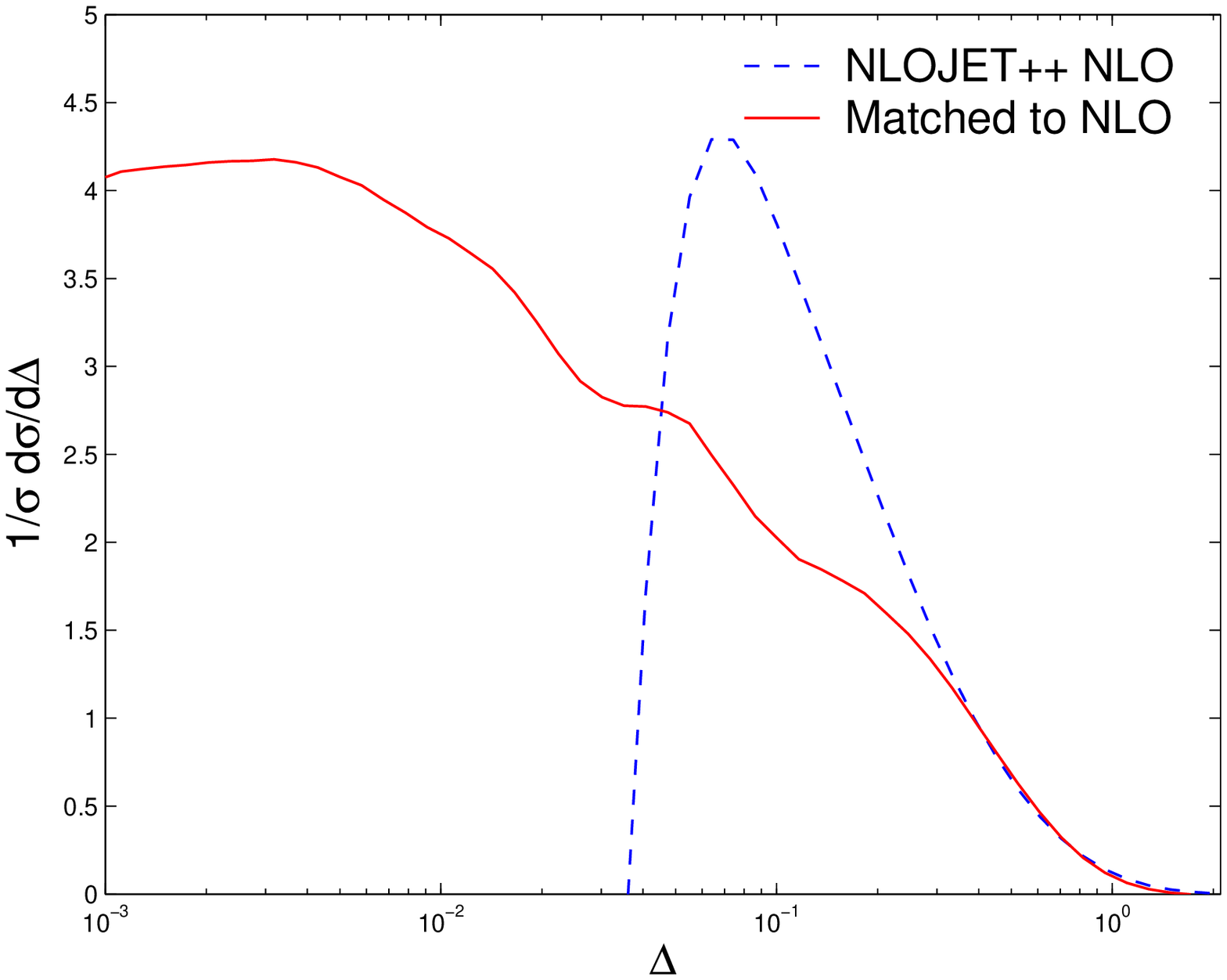,width=0.49\textwidth}\hfill
\epsfig{file=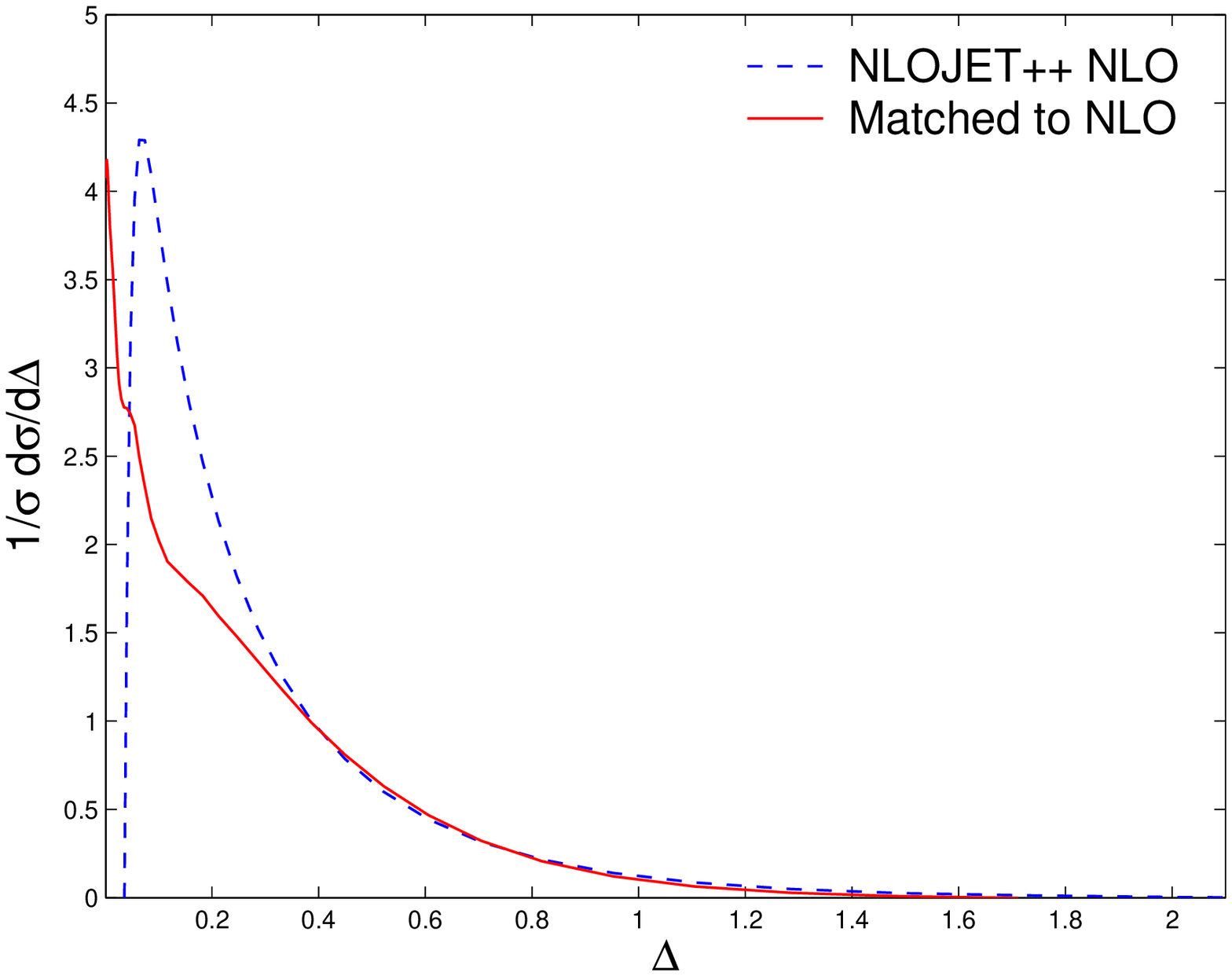,width=0.49\textwidth}
\caption{The Dijet azimuthal correlation distribution in DIS. A
comparison between the NLO and matched results is shown. The NLO
result diverges when $\Delta\rightarrow 0$, while the matched result
tends to a constant. The effect of matching is to bring the
distribution to NLOJET++ \cite{Nagy:2001xb} curve at large $\Delta$
(where we expect the NLO result to hold) and to correct the resummed
result at small $\Delta$ by a constant factor, not accounted for in
the resummation.} \label{Fig:delenda_fig}
\end{figure}

The resummed result needs to be corrected to include pieces which
are not captured by the resummation. Many techniques have been
developed to match resummed predictions with NLO results. Below we
report the matching formula we use here:
\begin{equation}
\sigma_{\mathrm{mat}} = \sigma(\Delta)
\left[1+\left(\sigma_{e}^{(1)}-\sigma_r^{(1)}\right)/
\sigma_0\right] +\left(\sigma_{e}^{(2)}-\sigma_r^{(2)}\right)
\exp\left(-R_{\mathrm{DL}}\right),
\end{equation}
where $\sigma(\Delta)$ is the resummed result, $\sigma_r^{(1)}$ and
$\sigma_r^{(2)}$ are the expansion of the resummed result to
$\mathcal{O}(\alpha_s)$ and $\mathcal{O}(\alpha_s^2)$ respectively
and $\sigma_0$ is the Born cross section. Here $\sigma_e$ denotes
the integrated distributions given by NLOJET++ \cite{Nagy:2001xb},
with the superscripts indicating the order, and $R_{\mathrm{DL}}$ is
the double logarithmic piece of the radiator obtained by replacing
$b\rightarrow
 e^{-\gamma_E}/\Delta$, where $\gamma_E$ is the Euler constant. We
present the results in Fig.~\ref{Fig:delenda_fig}.

\section{Hadronic collisions case}

We report below the result for the dijet azimuthal correlation
distribution in hadronic collisions. This has been measured at D\O\
using the jet recombination scheme in which the four-momentum of a
jet is obtained by the sum over the four-momenta of particles in the
jet. Here we only report the result which exploits Eq.
\eqref{eq:delenda_def} although similar results can be obtained for
the other scheme. The resummed result is given by:
\begin{equation}
\sigma(\Delta) = \int d\mathcal{B}
\frac{2}{\pi}\int_0^{\infty}\frac{db}{b} \sigma_{\mathcal{B}}(b)
\sin(b\Delta) e^{-R(b)}\times S\,,
\end{equation}
where
\begin{equation}
S=\mathrm{Tr} (H e^{-SL(b)\Gamma^{\dag}/2}M e^{-SL(b)\Gamma/2} )
/\mathrm{Tr} (H M)\,,
\end{equation}
with $H$, $\Gamma$ and $M$ being the hard, anomalous dimension and
soft matrices. These depend on the kinematics of the process and
appear in various places (e.g.
\cite{Botts:1989kf,Kidonakis:1996aq,Kidonakis:1998nf}). The single
logarithmic function $SL(b)$ accounts for soft wide-angle emissions.
Here $\sigma_{\mathcal{B}}(b)$ is the Born cross-section for the
production of two jets in hadronic collisions, which also contains
pdfs from both incoming legs evolved to $\mu_f^2/b^2$, and
$d\mathcal{B}$ is the corresponding phase-space for the production
of a dijet system in hadronic collisions. Note that the radiator in
this case has a slightly different form than in the DIS case.

\section{Future directions}

Having performed an NLL resummation (and NLO matching in the DIS
case) we can now compare our predictions with data and other
approaches (e.g. that of Ref. \cite{Jung} which implements
unintegrated pdfs).

We can further our study by looking at the hadronic collisions case
using the same jet definitions as those used by the D\O\
collaboration. The current indication is that the size of the
non-global component and the impact of the jet algorithm on the
``global'' piece may not be significant
\cite{Delenda:2006nf,Banfi:2003jj}, particularly since these pieces
contain only single logarithms while the distribution is dominated
by double logarithms.


\begin{footnotesize}



%

\end{footnotesize}


\end{document}